\documentclass[12pt,a4paper]{iopart}

\pdfoutput=1

\usepackage{iopams}
\usepackage[colorlinks=true, linkcolor=blue, citecolor=blue, urlcolor=blue]{hyperref}

\usepackage{graphicx}

\newcommand{\tem}{\ensuremath{\mathrm{TEM}_{00}}}
\newcommand{\erm}[1]{\ensuremath{E_{\mathrm{#1}}}}
\newcommand{\irm}[1]{\ensuremath{I_{\mathrm{#1}}}}
\newcommand{\prm}[1]{\ensuremath{P_{\mathrm{#1}}}}
\newcommand{\mvec}[1]{\bf{#1}}

\newcommand{\reffig}[1]{figure~\ref{#1}}
\newcommand{\refFig}[1]{Figure~\ref{#1}}

\begin{document}

\title{Cavity-Enhanced Rayleigh Scattering}

\author{Michael Motsch, Martin Zeppenfeld, Pepijn~W~H~Pinkse\footnote{Present address: Mesa+ Institute for Nanotechnology, University of Twente, PO Box 217, 7500 AE Enschede, The Netherlands} and Gerhard Rempe}
\address{Max-Planck-Institut f{\"u}r Quantenoptik, Hans-Kopfermann-Stra{\ss}e 1, 85748 Garching, Germany}
\ead{\mailto{gerhard.rempe@mpq.mpg.de}}

\begin{abstract}
We demonstrate Purcell-like enhancement of Rayleigh scattering into a single optical mode of a Fabry-Perot resonator for several thermal atomic and molecular gases. The light is detuned by more than an octave, in this case by hundreds of nanometers, from any optical transition, making particle excitation and spontaneous emission negligible. The enhancement of light scattering into the resonator is explained quantitatively as an interference effect of light waves emitted by a classical driven dipole oscillator. Applications of our method include the sensitive, non-destructive in-situ detection of ultracold molecules.
\end{abstract}

\maketitle

\section{Introduction}

Light scattering lies at the heart of optics. To harvest weak signals, collecting a large fraction of scattered light is essential. This is most directly achieved using a lens with a large numerical aperture. An alternative and more powerful approach, however, is to couple the scattering object to an optical resonator, even if the relevant light mode covers only a small solid angle. The potential offered by the resonator comes from the Purcell effect \cite{Purcell1946}, generally associated with the enhanced spontaneous emission rate of an excited particle in electromagnetically confined space.

Although spontaneous emission is a quantum-mechanical phenomenon, the modification of its rate by the resonator can be explained classically as a light interference effect \cite{Kastler1962, Milonni1973, Heinzen1987, Dowling1993, Friedler2009}. This makes the Purcell enhancement a universal phenomenon, independent of the internal structure of the particles under investigation. The Purcell effect therefore occurs both for light scattering from quantum objects with discrete energy levels and classical objects described as oscillating dipoles. While the Purcell effect has extensively been studied for quantum-mechanical objects over the last several decades, experiments involving classical objects are rare. Despite its conceptual simplicity, the effect of a cavity on the classical light-scattering properties has been scrutinized only recently using relatively complex systems such as subwavelength-sized fiber tips \cite{Mazzei2007} or silicon nanocrystals \cite{Kippenberg2009}. Moreover, the simplicity of the classical description has not been emphasized in \cite{Mazzei2007}, where the scattering of far-detuned light into the cavity mode was described by a semiquantum model for the interaction between the classical oscillating dipole and the quantized modes of the light field. Hence, this leaves open the question about the classical nature of the Purcell effect. Equally important, a demonstration of the Purcell effect for the detection of extremely weak signals in the far-detuned, classical regime is still lacking. The enhancement of such Rayleigh scattering from atomic or molecular gases by means of an optical resonator would enable one to observe in a sensitive and non-destructive way particles for which a closed two-level system or a near-resonant laser is not available.

Here we report on an experiment where the enhancement of Rayleigh scattering into an optical Fabry-Perot resonator is quantitatively studied using various thermal gases consisting of either atoms (Xe), homonuclear (N$_2$) or heteronuclear molecules (CF$_3$H). The light is detuned by hundreds of nanometers from the nearest electronic or vibrational transition. This detuning by more than one octave makes excitation of the system negligible. For different values of the cavity finesse, we observe an increased rate of Rayleigh scattering, in agreement with the predictions of a completely classical model based on the interference of intracavity light waves. Firstly, this proves that the Purcell effect is classical, and secondly, it shows that an explanation in terms of a modified local mode density is not needed. Comparing the power scattered into the fundamental cavity mode to that scattered into the same mode but under free-space conditions, we find an enhancement by a factor of up to 38.

\section{Experimental Setup}

\begin{figure}
\begin{indented}
\item[]
\includegraphics[scale=1.]{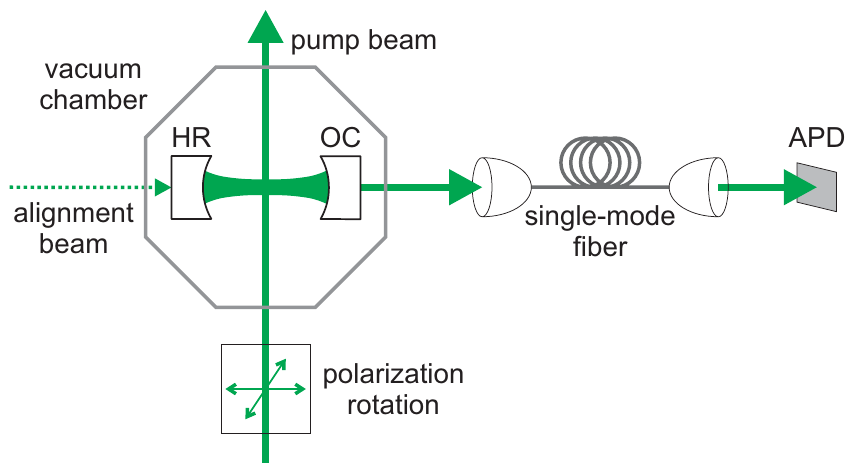}
\caption{The experimental setup. The cavity consists of two mirrors, a high reflector (HR) and an out coupler (OC), placed in a vacuum chamber, which can be filled with various gases. The polarization direction of the linearly polarized pump beam is rotated by a Pockels cell. The beam waist is adjusted to match the waist of the fundamental cavity mode \tem. Scattered light leaking out of the cavity through the OC is mode matched into a single-mode fiber, and then detected by an avalanche photodiode (APD).}
\label{fig:ExpSetup}
\end{indented}
\end{figure}

The experimental setup is depicted in \reffig{fig:ExpSetup}. A 10\,W single-frequency laser at wavelength $\lambda$=532\,nm was used. This wavelength was chosen as a compromise between the $\lambda^{-4}$ dependence of the Rayleigh scattering cross section on the laser wavelength, available laser power and sufficient detuning from deep-ultraviolet electronic transitions of the used species. The cavity mirrors are separated by 0.6\,cm, and the mirrors' radius of curvature is 4.5\,cm. This combination results in a spacing of transverse modes of 4.1\,GHz, larger than the observed Doppler widths of the gases used. The cavity length is tuned by a piezoelectric tube separating the two mirrors. To modify the cavity finesse, the mirror on the outcoupling side (OC) is exchanged, while the other mirror (HR, $\mathrm{R}$=99.7\,\%) is used for all measurements. For the various combinations of mirrors with measured intensity reflectivities $\mathrm{R}$=(99.7\,\%, 98.9\,\%, 95.9\,\%), a cavity finesse of  $\mathcal{F}$=(1000, 400, 100) is determined from the cavity line width as observed in transmission. The cavity is placed inside a vacuum chamber, which is pumped out to $10^{-2}$\,mbar before the various gases are introduced. The pump beam was focussed to a waist of about 50\,$\mu\mathrm{m}$ as measured by a beam profiler. This value was chosen to match the waist of the fundamental cavity mode, which was calculated to be $w_0=45\,\mu\mathrm{m}$. Since we pump with powers of up to 5W and expect a scattered power into the cavity mode of only a few fW, suppression of stray light is crucial. Towards this end, light leaving the cavity through the outcoupling mirror is send to a single-mode fiber which is aligned for optimum transmission of the {\tem} fundamental cavity mode. Behind the fiber, the light is detected by an avalanche photodiode operated in single-photon-counting mode.

\section{Frequency Dependence of the Rayleigh-Scattered Light}

\begin{figure}
\begin{indented}
\item[]
\includegraphics[scale=1.]{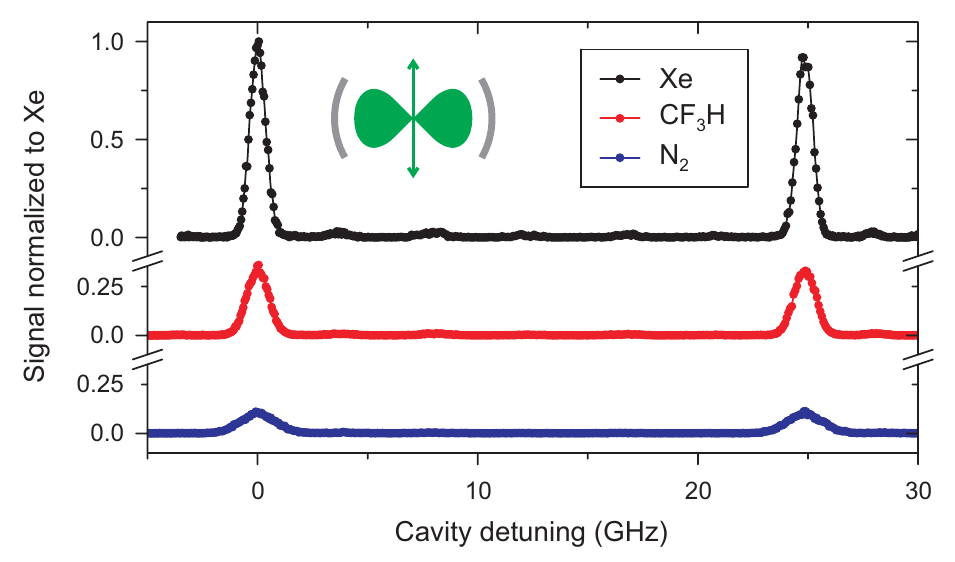}
\caption{
Cavity output measured behind the single-mode fiber. The resonances of the fundamental transverse {\tem} mode are separated by the cavity's free spectral range of 24.9\,GHz. The signal of the individual gases is normalized to the signal obtained with Xe. The line profile originates from the Doppler broadening of the light scattered by the thermal gases.
}
\label{PhD:fig:rayleigh:ModeSpec}
\end{indented}
\end{figure}

In the experiment, we align the pump beam perpendicular to the cavity axis as indicated in \reffig{fig:ExpSetup}, scan the cavity over at least one free-spectral range and monitor the intensity of the light leaking out of the cavity. As shown in \reffig{PhD:fig:rayleigh:ModeSpec}, we observe mainly light scattered into the {\tem} mode, which is selected by the single-mode fiber. A small amount of light scattered into other cavity modes is visible as well due to imperfect mode matching into the fiber. For the various gases used, the peaks show different heights and widths. The different widths originate from the different masses and, hence, the different Doppler broadenings of the thermal gases. The different heights are caused by the specific polarizabilities of the different particles as well as by the frequency overlaps of the cavity mode with the Doppler profiles. Within experimental accuracy, the background signal observed for N$_2$ and CF$_3$H agrees with the one obtained for Xe. Doppler broadening and the multitude of thermally populated internal molecular states prevent the observation of a Raman spectrum, which could benefit from a cavity enhancement as well \cite{Morigi2007}.

\begin{figure}
\begin{indented}
\item[]
\includegraphics[scale=1.]{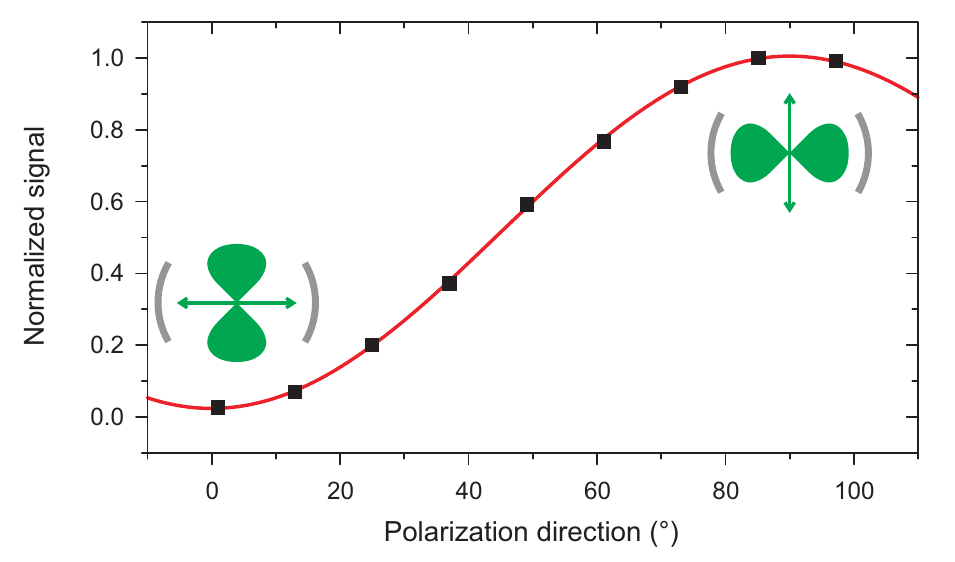}
\caption{
Measured polarization dependence of the Rayleigh-scattered light from any of the three gases used. When the polarization direction of the pump beam is aligned with the cavity axis, no scattering into the cavity occurs. In the opposite case, the polarization direction being perpendicular to the cavity axis, a maximal amount of light is scattered into the cavity. The solid curve is a fit of a $\sin^2\theta$ polarization dependence to the data. For light scattering from a classical oscillating dipole, such a $\sin^2\theta$ polarization dependence is expected.
}
\label{PhD:fig:rayleigh:PolDep}
\end{indented}
\end{figure}

A signature of Rayleigh scattering is the $\sin^2\theta$ polarization dependence when driven by linearly polarized light. Here, $\theta$ denotes the azimuthal angle of the pump beam's polarization direction with respect to the cavity axis. \refFig{PhD:fig:rayleigh:PolDep} shows the polarization dependence of light scattered into the \tem\ mode of the cavity, when $\theta$ is rotated using a Pockels cell. It follows a $\sin^2\theta$ dependence, as expected for emission from a classical oscillating dipole. When the polarization direction is aligned with the cavity axis $(\theta=0^\circ)$, no Rayleigh scattering into the cavity mode is expected due to the small solid angle of the cavity covering only the nodal line of the dipole pattern. The residual signal amplitude of 1--2\,\% can be explained by imperfect linear polarization of the pump beam. We emphasize that the $\sin^2\theta$ dependence, observed for all the gases investigated, is the characteristic feature for a scattering process involving an oscillating dipole.

\begin{figure}
\begin{indented}
\item[]
\includegraphics[scale=1.]{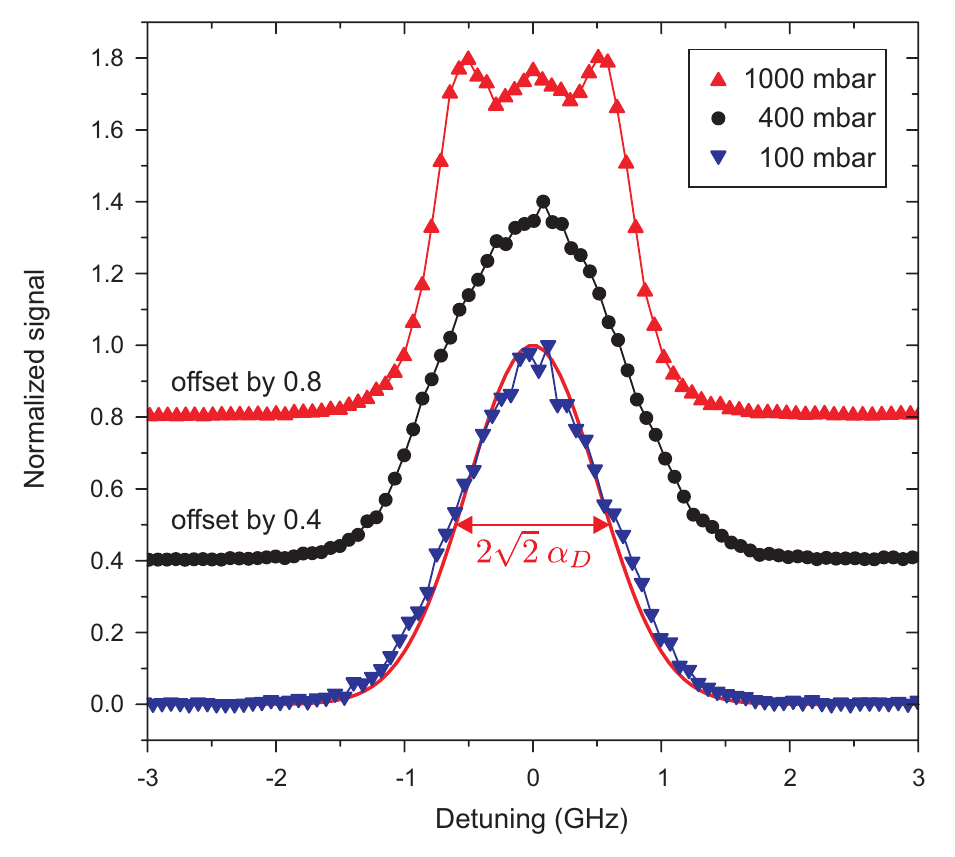}
\caption{
Doppler-broadened line profiles for CF$_3$H\@. The data of the cavity scans obtained for the different pressures have been individually normalized to the maximal value. In the measurement with the lowest pressure, the theoretically predicted Doppler profile is shown in red. At higher pressure, sidebands caused by Brillouin scattering appear.
}
\label{PhD:fig:rayleigh:PressDep}
\end{indented}
\end{figure}

At low densities, the line shape is determined by Doppler broadening as shown in \reffig{PhD:fig:rayleigh:PressDep}. Due to the $90^\circ$ scattering geometry, we observe a Doppler width $\alpha_\mathrm{obs}$=$\sqrt{2}\;\alpha_{D}$, with $\alpha_{D}$ being the standard Doppler width of absorption spectroscopy. By varying the pressure, we observe a linear dependence of the scattered power on the gas density, thereby ruling out scattering off the surfaces as the signal source. We also find the expected linear dependence of the Rayleigh scattered light on the pump beam power over the entire examined range of 0.1--5\,W. However, when increasing the gas pressure from 100\,mbar, where we typically operate, to values of up to 1\,bar, we observe the appearance of a substructure on the Doppler-broadened peaks. These sidebands are caused by Brillouin scattering on density waves in the gas \cite{Ghaem1980, Pan2002}.

\section{Classical Wave Interference Model of Cavity Enhancement}
\begin{figure}
\begin{indented}
\item[]
\includegraphics[width=0.83\columnwidth]{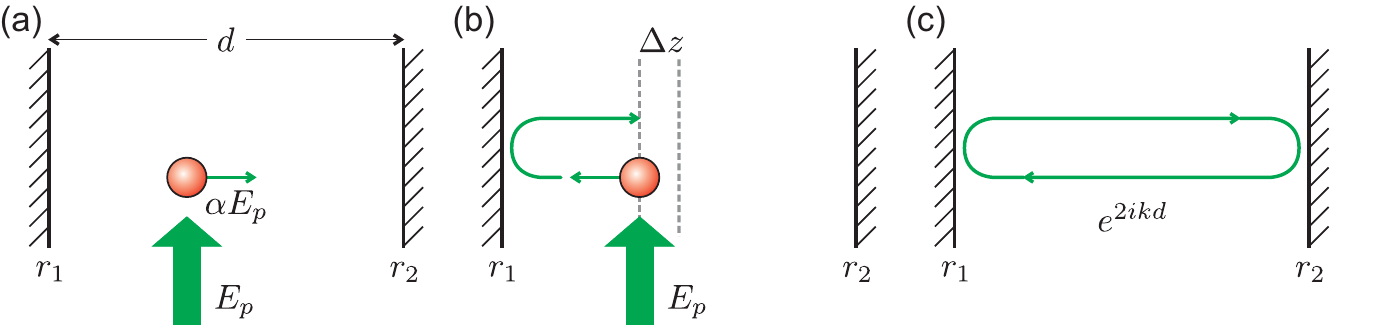}
\caption{
Theory sketch. Different contributions to the right-traveling intracavity field at the position of the scattering particle. (a) Field scattered into the right direction. (b) Field scattered to the left, after one reflection at the mirror. (c) Intracavity field after one round trip.
}
\label{fig:rayleigh:Theory}
\end{indented}
\end{figure}

To model the experiment, we describe the scattering from a polarizable particle, a classical oscillating dipole, into a single cavity mode as follows. The polarizable particle is pumped from the side by the electric field \erm{p} and scatters an electric field into the cavity as indicated in \reffig{fig:rayleigh:Theory}. To derive an expression for the intracavity light field and the power scattered into the cavity mode, we follow the intracavity light field on one round trip through the cavity. Along its way, the intracavity light field grows due to light scattering from the pump beam into the cavity mode and decreases due to transmission losses through the mirrors.

As schematically shown in \reffig{fig:rayleigh:Theory}, the scatterer emits an electromagnetic field into a right-traveling and a left-traveling wave, which can interfere. The right-running intracavity field $\erm{c}$ at the position of the scatterer is
\begin{equation}
\label{eq:CavityField}
\erm{c} = \alpha \erm{p} + \mathrm{r}_1 e^{ik(d+2\Delta z)} \alpha \erm{p} + \mathrm{r}_1 \mathrm{r}_2 e^{2ikd} \erm{c}.
\end{equation}
Here, $\alpha$ is a proportionality factor for the scattering process, and $\mathrm{r}_i$ $(i=1,2)$ are the mirror's amplitude reflection coefficients ($\mathrm{r}_i^2=\mathrm{R}_i$ is the mirror reflectivity). The distance between the mirrors is denoted by $d$, and $k=2\pi/\lambda$ is the wave number of the light field. The first term of \eref{eq:CavityField}, $\alpha \erm{p}$, is the field directly scattered into the right direction. The second term of \eref{eq:CavityField}, $\mathrm{r}_1 e^{ik(d+2\Delta z)} \alpha \erm{p}$, is the field initially scattered into the left direction. After being reflected by the left cavity mirror, it arrives at the position of the scatterer as a right-traveling electric field. The additional phase factor $e^{2ik\Delta z}$ accounts for a translation $\Delta z$ of the scatterer from the cavity centre. The last term of \eref{eq:CavityField}, $\mathrm{r}_1 \mathrm{r}_2 e^{2ikd} \erm{c}$, describes the right-traveling intracavity field after one round trip.
By solving \eref{eq:CavityField}, one finds for the right-traveling intracavity field
\begin{equation}
\label{eq:CavityField2}
\erm{c} = \alpha \erm{p} \, \frac{1+\mathrm{r}_1\,e^{ikd} \, e^{2ik\Delta z}}{1-\mathrm{r}_1 \mathrm{r}_2\,e^{2ikd}}.
\end{equation}
In the following discussion, we assume a resonant cavity, $e^{2ikd}=1$. As a first step, we use \eref{eq:CavityField2} to calculate the right-traveling intensity,
\begin{math}
\irm{c} = \frac{c\epsilon_0}{2}\left|\erm{c}\erm{c}^*\right|.
\end{math}
Since the experiment is performed with thermal gases, the scattering particles are randomly distributed. Therefore, an average over the position of the particles, i.e., $-\lambda/2 \leq \Delta z \leq \lambda/2$, must be taken, resulting in
\begin{equation}
\label{eq:runningIntensity}
\irm{c}  = \alpha^2 \irm{p} \, \frac{1+\mathrm{r}_1^2}{\left(1-\mathrm{r}_1 \mathrm{r}_2\right)^2}.
\end{equation}
In the limit of high mirror reflectivities, $\mathrm{R}_i=\mathrm{r}_i^2 \approx 1$, the right-traveling intracavity intensity \eref{eq:runningIntensity} can be expressed in a form with a clear physical meaning. With the cavity finesse $\mathcal{F}=\pi\sqrt[4]{\mathrm{R}_1\mathrm{R}_2}/(1-\sqrt{\mathrm{R}_1\mathrm{R}_2}) \approx\pi/(1-\sqrt{\mathrm{R}_1\mathrm{R}_2})$ one finds
\begin{equation}
\label{eq:runningIntensity2}
\irm{c} \approx 2\alpha^2 \irm{p} \left(\frac{\mathcal{F}}{\pi}\right)^2.
\label{eq:Ic}
\end{equation}
In an intuitive picture, $\mathcal{F}/\pi$ is the number of reflections in the resonator. The right-traveling electric field increases linearly with this number. The intensity is proportional to the square modulus of the electric field, which results in $\irm{c} \propto(\mathcal{F}/\pi)^2$.

As a next step, we quantify the enhancement of the power emitted from the cavity. We compare the power of the scattered light leaving the resonator through one of the cavity mirrors to the power of the light scattered into the same mode but under free-space conditions. For this and the following discussions, it is convenient to convert all intensities $I$ into powers $P$ taking the area of the cavity mode as a reference. From \eref{eq:Ic}, the power leaving the cavity through the right mirror is found to be
\begin{equation}
\prm{t} = \mathrm{T}_2 \times \prm{c} \approx 4 \, \frac{\mathrm{T}_2}{\mathrm{T}_1+\mathrm{T}_2} \, \alpha^2\prm{p} \; \frac{\mathcal{F}}{\pi},
\end{equation}
when the Taylor series expansion
\begin{math}
\mathcal{F}/\pi \approx \left (1-\sqrt{\mathrm{R}_1\mathrm{R}_2}\right)^{-1} \approx 2 \left(\mathrm{T}_1+\mathrm{T}_2\right)^{-1}
\end{math}
and $\mathrm{r}=\sqrt{\mathrm{R}}=\sqrt{1-\mathrm{T}}$ are used. In the case of a symmetric cavity, $\mathrm{T}_1=\mathrm{T}_2=\mathrm{T}$, this can be simplified to yield
\begin{equation}
\label{eq:TransPow}
\prm{t} \approx 2 \, \alpha^2 \prm{p}\; \frac{\mathcal{F}}{\pi}.
\end{equation}
Without the cavity in place, the right-traveling scattered field is given by \begin{math}\erm{rt}=\alpha \erm{p}\end{math}. The power scattered into the right-traveling mode defined by the cavity but under free-space conditions is therefore given by \begin{math} \prm{rt}=\alpha^2 \prm{p}\end{math}. Comparing this with the power leaving the cavity through the right mirror, \eref{eq:TransPow}, the cavity enhances the detectable power by a factor $2\;\mathcal{F}/\pi$.

Half of the light, which has been scattered into the cavity mode, leaks out of the left cavity mirror. Taking this into account, the total power scattered into the cavity mode is given by
\begin{equation}
\prm{cav} = 2\,\prm{t} \approx4\,\alpha^2\prm{p}\; \frac{\mathcal{F}}{\pi}.
\end{equation}
For particles maximally coupled to the cavity, i.e., no averaging over particle position, an additional factor 2 comes in,
\begin{equation}
\prm{cav} = 8 \,\alpha^2\prm{p} \; \frac{\mathcal{F}}{\pi}.
\end{equation}
This has to be compared to the total power scattered under free-space conditions into the mode $\Omega_{\mathrm{cav}}$ defined by the cavity but without the cavity enhancement, which is
\begin{equation}
\label{eq:PowerFreeSpaceCavityMode}
\prm{fs}^{\Omega_{\mathrm{cav}}} =  2 \, \prm{rt} = 2 \, \alpha^2 \prm{p}.
\end{equation}
Therefore, the cavity enhances the power scattered into the same mode of the electromagnetic field by a factor
\begin{math}
4\,\mathcal{F}/\pi
\end{math}
as compared with the free-space situation.

So far, only a single mode of the electromagnetic field was considered. Thereby, the power scattered into the cavity mode was compared to the power scattered into the same mode in the free-space situation. Such a scenario refers to our experiment. It is, however, also interesting to compare the power $\prm{cav}$ scattered into the cavity mode with the power $\prm{fs}$ scattered into the full $4\pi$ solid angle under free-space conditions. In the end, this calculation allows to establish a connection between the classical wave interference model and the well-known Purcell factor \cite{Purcell1946}.

\begin{figure}
\begin{indented}
\item[]
\includegraphics[width=.83\columnwidth]{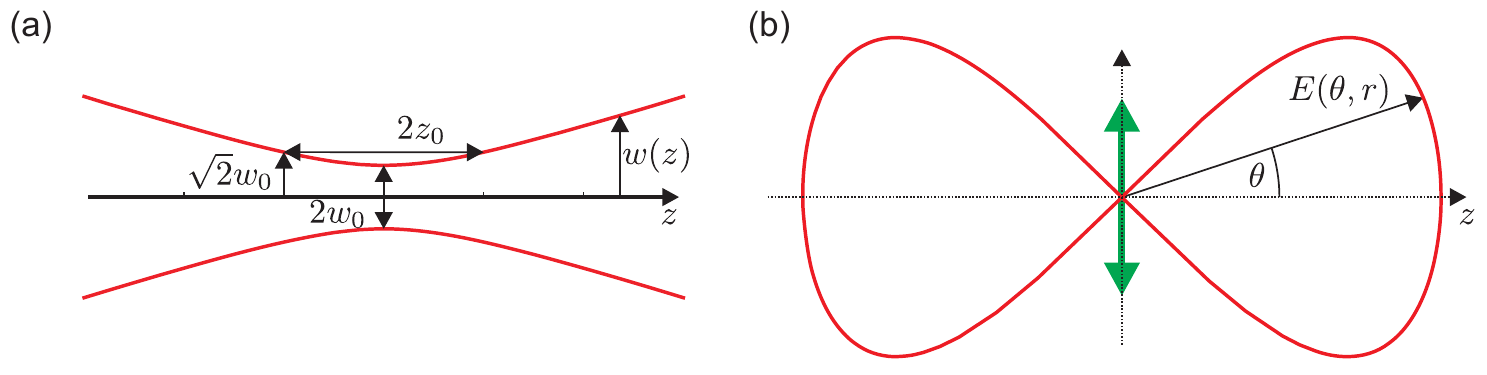}
\caption{
(a) Beam width $w(z)$ of the lowest-order transverse Hermite-Gaussian cavity mode {\tem} along its propagation direction $z$. $w_0$ is the beam waist, and $z_0=\pi w_0^2/\lambda$ is the Rayleigh length.
(b) Emission pattern of the classical oscillating dipole oriented perpendicular to the $z$-axis.
}
\label{fig:rayleigh:Modes}
\end{indented}
\end{figure}

As a starting point, we calculate which fraction of the light scattered by the oscillating dipole is radiated into the mode defined by the cavity but under free-space conditions. For this purpose, the overlap integral between the intensity-normalized electric fields of the dipole mode and the cavity mode are evaluated. The cavity mode and the dipole emission pattern are schematically shown in \reffig{fig:rayleigh:Modes}. The oscillating dipole is oriented perpendicular to the cavity axis, which results in maximal light scattering into the cavity mode. In the far field, the scalar dipole mode $\erm{dip}(\theta,r)$ can be defined in spherical coordinates as
\begin{equation}
\label{eq:DipMode}
\erm{dip}(\theta,r)=\sqrt{\frac{3}{8\pi}}\frac{1}{r}\cos(\theta), \hspace{1cm} (-\frac{\pi}{2}\leq\theta\leq\frac{\pi}{2}),
\end{equation}
where the phase variation $\exp(ikr)$ along the propagation direction $\mvec{r}$ has been omitted. The dipole mode is normalized to its intensity,
\begin{equation}
\int_0^{2\pi} \int_{-\frac{\pi}{2}}^{\frac{\pi}{2}} \erm{dip}(\theta,r)^2 \; r^2 \cos(\theta) \mathrm d\theta \mathrm d\varphi = 1.
\end{equation}
In the experiment, the lowest-order transverse Hermite-Gaussian mode {\tem} was used. The electric field of the fundamental cavity mode $\erm{cav}(R,z)$ in cylindrical coordinates is given by
\begin{equation}
\label{eq:CavMode}
\erm{cav}(R,z) = \frac{1}{N_{\mathrm{cav}}(z)}\exp{\left(-\frac{R^2}{w(z)^2}\right)}.
\end{equation}
Again, all phase variations have been omitted, as justified below. $N_{\mathrm{cav}}(z)$ is a normalization constant, and $w(z)$ is the beam waist.
This cavity mode traveling in one direction is normalized in cylindrical coordinates,
\begin{equation}
\int_{0}^{2\pi} \int_0^\infty \erm{cav}(R,z)^2 \; R \, \mathrm dR \mathrm d\varphi=1.
\end{equation}

Now, the overlap integral between the intensity-normalized electric fields of the dipole mode $\erm{dip}$, \eref{eq:DipMode}, and the fundamental Hermite-Gaussian cavity mode $\erm{cav}$, \eref{eq:CavMode}, is evaluated in the far field at a fixed value $z$. There, both the dipole mode and the cavity mode have phase fronts lying on spheres centered at the position of the scatterer. All phase factors therefore cancel in the evaluation of the overlap integral between the two modes, which justifies the initial disregard of these. For the real-valued dipole and cavity modes, the overlap in one propagation direction $z$ is given by
\begin{equation}
\label{eq:OverlapInt}
\eta = \int_A \erm{dip} \erm{cav} \mathrm d\Omega,
\end{equation}
which is evaluated in cylindrical coordinates. Due to the small transverse extent of the cavity mode, the dipole mode $\erm{dip}$ is approximated by its value on the $z$-axis,
\begin{math}
\erm{dip}(\theta=0,z)=\sqrt{3/8\pi}\,z^{-1}.
\end{math}
In the limit $z \rightarrow \infty$, this results in
\begin{equation}
\eta = \frac{\sqrt{3}}{2\pi} \frac{\lambda}{w_0}.
\end{equation}
Using \eref{eq:OverlapInt}, the overlap between the electric fields of the dipole mode and the cavity mode was evaluated. Now, the power scattered into the cavity mode $\prm{fs}^{\Omega_{\mathrm{cav}}}$ and the total power scattered into the dipole mode $\prm{dip}$ are compared. Since power scales with the square of the electric field, one finds
\begin{equation}
\label{eq:FracPowerCavMode}
\frac{\prm{fs}^{\Omega_{\mathrm{cav}}}}{\prm{dip}} = 2\eta^2 = \frac{3}{2\pi^2} \frac{\lambda^2}{w_0^2}.
\end{equation}
In the evaluation of the overlap integral \eref{eq:OverlapInt} only one propagation direction of the cavity mode was taken into account. The factor 2 in \eref{eq:FracPowerCavMode} accounts for the two independent propagation directions of the cavity mode. Since there is no interference between the fields scattered into these two directions, intensities are summed up, not fields. Using \eref{eq:PowerFreeSpaceCavityMode} and \eref{eq:FracPowerCavMode}, the total power scattered into the dipole mode is found to be
\begin{equation}
\label{eq:PowerDipoleMode}
\prm{dip} = \frac{\prm{dip}}{\prm{fs}^{\Omega_{\mathrm{cav}}}} \times \prm{fs}^{\Omega_{\mathrm{cav}}}
= \frac{4\pi^2 w_0^2}{3\lambda^2} \, \alpha^2 \prm{p}.
\end{equation}

To retrieve the Purcell factor, the power $\prm{cav}$ scattered into the cavity mode by a particle maximally coupled to the cavity is compared with the power $\prm{dip}$ scattered by the particle into the dipole mode under free-space conditions. One finds
\begin{equation}
\label{eq:PowerRatio}
\frac{\prm{cav}}{\prm{dip}}
= \frac{8 \, \alpha^2 \prm{p} \; {\displaystyle \frac{\mathcal{F}}{\pi}}}
{ {\displaystyle \frac{4\pi^2 w_0^2}{3\lambda^2}} \, \alpha^2 \prm{p}}
= \frac{6}{\pi^2}\;\frac{\lambda^2}{w_0^2} \; \frac{\mathcal{F}}{\pi}.
\end{equation}

In its most common form, the Purcell factor \cite{Purcell1946} is defined as
\begin{equation}
2C = \frac{3}{4\pi^2} \; Q \; \frac{\lambda^3}{V},
\end{equation}
with $Q$ being the quality factor of the cavity and $V$ the mode volume. For an atomic system, the free-space decay rate $\Gamma$ of an excited state is changed to the value $\Gamma'=(1+2C)\,\Gamma$ by the presence of a cavity being resonant with the atomic transition frequency. For a two-mirror Fabry-Perot cavity with mirror separation $d$, the mode volume $V$ is well approximated by
\begin{math}
V=\pi w_0^2 d/4.
\end{math}
The $Q$ factor is defined as $Q=\nu/\delta\nu$ with $\nu$ being the resonance frequency of the cavity and $\delta\nu$ the cavity linewidth. For a two-mirror Fabry-Perot cavity, this can be related to the cavity finesse $\mathcal{F}$ and the mirror separation $d$ by
\begin{math}
\mathcal{F} = c / (2d\,\delta\nu)
\end{math}
and
\begin{math}
Q=\nu/\delta\nu=c/(\lambda\,\delta\nu).
\end{math}
Therefore, the cavity $Q$ factor can be expressed in terms of the cavity finesse $\mathcal{F}$ and mirror separation $d$ as
\begin{math}
Q=2d\,\mathcal{F}/\lambda.
\end{math}
Putting all of the above together, one finds for the Purcell factor of a standing-wave Fabry-Perot cavity at its antinode
\begin{equation}
2C= \frac{3}{4\pi^2} \, Q \; \frac{\lambda^3}{V}
= \frac{6}{\pi^2}\;\frac{\lambda^2}{w_0^2} \; \frac{\mathcal{F}}{\pi}.
\end{equation}
This is exactly the same expression as the one derived from the classical wave-interference model \eref{eq:PowerRatio}. This shows that the Purcell factor is indeed fully explained by interference of scattered fields \cite{Kastler1962, Milonni1973, Heinzen1987, Dowling1993, Friedler2009}.

\section{Cavity-Finesse Dependence of Rayleigh Scattering}

\begin{figure}
\begin{indented}
\item[]
\includegraphics[scale=1.]{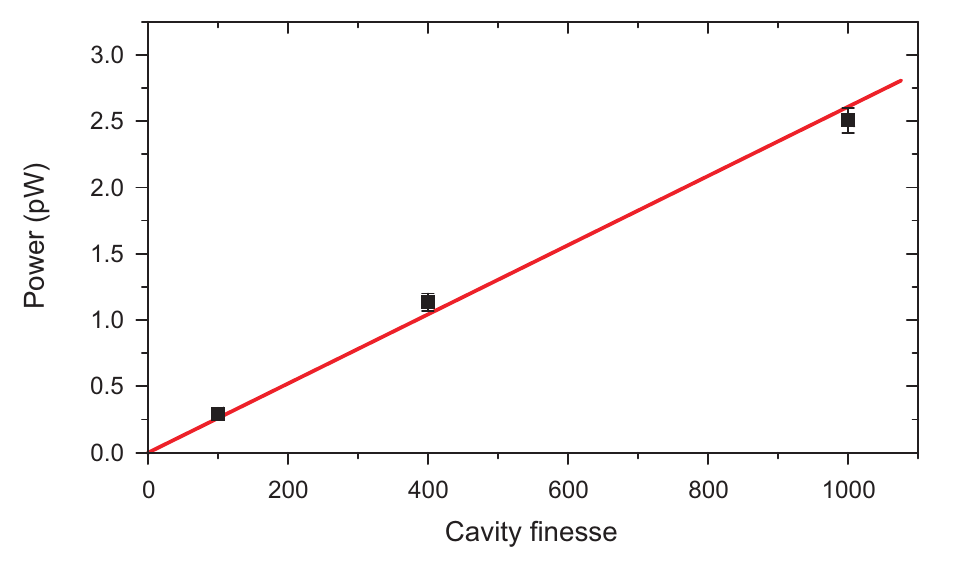}
\caption{
Dependence of scattered power on the cavity finesse, measured with Xe at 0.1\,bar. The data are corrected for the limited overlap of the cavity mode with the Doppler-broadened spectrum of the Rayleigh-scattered light.
}
\label{PhD:fig:rayleigh:FinesseDep}
\end{indented}
\end{figure}

To test the expected dependence of the scattered power on the cavity finesse, we measured the cavity output power for the three available values $\mathcal{F}$=(1000,\,400,\,100). The spectral overlaps of a Doppler-broadened thermal Xe gas with the cavity modes are (4.2,\,10.1,\,33.4)\,\% for the used cavities, respectively. From the measured signals $(52\pm2,\,85\pm5,\,90\pm5)\,\mathrm{fW}$, we calculate the power scattered from particles at rest and plot it in \reffig{PhD:fig:rayleigh:FinesseDep}. As predicted by the classical wave-interference model and by the Purcell theory, a linear dependence on the cavity finesse is found. The scattered power also depends on the polarizability of the particles. For fixed cavity finesse, we find a ratio between scattered powers for Xe, CF$_3$H, and N$_2$ of (1:\,0.35:\,0.1). Based on their  polarizabilities~\cite{HandbookChemPhys71st, Miller1981} and Doppler broadenings we expect a ratio of (1:\,0.36:\,0.09), which is in good agreement with the observations.

Finally, we show the enhancement of the scattered power leaving the cavity as compared with the free-space situation. For this measurement, the single-mode fiber was used to select the same mode in a free-space scattering experiment. Under otherwise identical conditions (100\,mbar Xe pressure, 1\,W pump power), but without the cavity in place, we observe a scattered power of $\approx1.3\,\mathrm{fW}$. We estimate a factor 2 for the accuracy of this measurement, constrained by mode-matching. This power measured in free-space scattering has to be compared with the value of $\approx50\,\mathrm{fW}$ measured with a cavity finesse of 1000, an enhancement factor of 38. To test the classical model for the cavity enhancement we take into account the limited overlap of the Doppler-broadened spectral profile with the cavity mode of 4.2\,\%, which does not occur in the free-space scattering, and the enhancement by the cavity with a finesse of 1000. Putting things together, we expect a value of $\approx2.0\,\mathrm{fW}$ for the free-space scattering measurement without the cavity, taking $\approx50\,\mathrm{fW}$ measured with the cavity as a reference. Within the experimental uncertainties, the calculated and measured enhancements are in agreement, showing that the process is well described by the classical model.

\section{Summary and Outlook}

The experiment shows the potential of an optical cavity to enhance weak signals in light-scattering experiments. The large detuning of the laser light from any optical transition allows the method to be applied to different species as demonstrated with the use of Xe, N$_2$ and CF$_3$H, independent of their specific internal level structure. This opens a new pathway to optical detection of deeply-bound ultracold molecules, for which the efficient production was reported recently \cite{Sage2005, Ospelkaus2008, Danzl2008, Deiglmayr2008, Lang2008, Viteau2008}. So far, however, detection of these molecules requires either dissociation to unbound atom pairs or ionization. Here, cavity-enhanced Rayleigh scattering might be an attractive in-situ detection technique since it does not rely on closed cycling transitions. Although in our experiment room-temperature gases at typical densities of $2\times10^{18}$\,cm$^{-3}$ were used, the number of particles contributing to the scattering into the cavity is only about $10^{10}$ for a cavity finesse of 1000.
The aforementioned molecule production techniques can typically prepare $10^5$ ultracold molecules at cloud sizes compatible with the cavity-mode diameter in the presented experiment. Since Doppler broadening is absent for light scattered by these trapped ultracold molecules, the cavity finesse could be increased to an experimentally realistic value of $10^5$. Furthermore, alkali dimers have a more than 10$\times$ larger polarizability $\alpha$ as compared with Xe (static polarizabilities $\alpha$, which should give a lower bound for the polarizabilities at optical frequencies: KRb: 502\,a.u., Cs$_2$: 670\,a.u., Xe: 27.3\,a.u., 1\,a.u.=0.148\,\AA$^3$) \cite{HandbookChemPhys71st, Deiglmayr2008a}, and the Rayleigh-scattering cross section $\sigma$ scales proportionally to $\alpha^2$. Altogether, ensemble scattering rates into the cavity of the order of 100\,kHz can therefore be anticipated for such an ultracold molecular sample. Hence, on average, every molecule scatters photons at a total rate of the order of 1\,Hz, with a ratio between scattering into the cavity and scattering into free space $P_{\mathrm{cav}}/P_{\mathrm{fs}}\approx2$ for the given parameters. This allows for a nearly demolition-free detection with almost no photon-recoil heating.
As an extension of the present setup, the use of cavities with degenerate transverse modes could boost the total power scattered into the cavity even further \cite{Chan2003}. Collective enhancement effects \cite{Domokos2002}, observed for atomic ensembles \cite{Black2003}, could result in an additional increase of signal.

\ack
The authors thank T.J. Kippenberg, M. Kowalewski and K. Murr for fruitful discussions. The support of the DFG through the excellence cluster "Munich Centre for Advanced Photonics" and EuroQUAM (Cavity-Mediated Molecular Cooling) is acknowledged.

\section*{References}

\bibliographystyle{unsrt}

\end{document}